\documentclass[preprint,12pt]{elsarticle}

\usepackage{amssymb}
\usepackage{amsmath}
\usepackage{amsthm}
\usepackage{verbatim}
\usepackage{makecell}
\usepackage{siunitx}
\usepackage{multirow}
\usepackage[caption=false,font=normalsize,labelfont=sf,textfont=sf]{subfig}
\usepackage{array}
\usepackage{textcomp}
\usepackage{stfloats}
\usepackage{url}
\usepackage{placeins}
\usepackage{paralist}
\usepackage{enumitem}
\usepackage{acro}
\usepackage{derivative}
\usepackage{eurosym}

\usepackage{notation}
\usepackage{acronyms}

\usepackage[noend, ruled, linesnumbered]{algorithm2e}

\graphicspath{{figures/}}

\newtheorem*{propositionrestated}{Proposition}
\SetKwInput{KwData}{Input}
\SetKwInput{KwResult}{Output}

\journal{Applied Energy}

\begin{document}

\begin{frontmatter}


\title{Optimal Intraday Power Trading for Single-Price Balancing Markets: An Adaptive Risk-Averse Strategy using Mixture Models}

\author[1]{Robin Bruneel\corref{cor1}}
\author[1]{Mathijs Schuurmans}
\author[1]{Panagiotis Patrinos}

\affiliation[1]{organization={Department of Electrical Engineering (ESAT-STADIUS), KU Leuven},
            addressline={Kasteelpark Arenberg 10},
            city={Leuven},
            postcode={3001}, 
            state={Vlaams-Brabant},
            country={Belgium}}

\cortext[cor1]{Corresponding author. Email: robin.bruneel@kuleuven.be}

\begin{abstract}
  Efficient markets are characterised by profit-driven participants continuously refining their positions towards the latest insights.
  Margins for profit generation are generally small, shaping a difficult landscape for automated trading strategies. 
  This paper introduces a novel intraday power trading strategy tailored for single-price balancing markets.
  The strategy relies on a strategically devised mixture model to forecast future system imbalance prices and is formulated as a stochastic optimization problem with decision-dependent distributions to address two primary challenges: (i) the impact of trading positions on the system imbalance price and (ii) the uncertainty inherent in the model.
  The first challenge is tackled by adjusting the model to account for price changes after taking a position. For the second challenge, a coherent risk measure is added to the cost function to take additional uncertainties into account. 
  This paper introduces a methodology to select the tuning parameter of this risk measure adaptively by continuously quantifying the performance of the strategy on 
  a window of recently observed data.
  The strategy is validated with a simulation on the Belgian electricity market using real-time market data. 
  The adaptive tuning approach leads to higher absolute profits, while also reducing the number of trades.
\end{abstract}



\begin{keyword}

Balancing markets \sep Electricity trading \sep  Mixture models \sep  Risk management \sep Stochastic optimization

\end{keyword}

\end{frontmatter}

\section*{Nomenclature}
\noindent \textit{Abbreviations} \\
MDP: marginal decremental price\\
MIP: marginal incremental price\\
BRP: balance responsible party\\
TSO: transmission system operator\\
SI: system imbalance\\
aFRR: automatic frequency restoration reserve\\
mFRR: manual frequency restoration reserve\\
CVaR: conditional value at risk\\
EVaR: entropic value at risk\\
pdf: probability density function\\\\
\textit{Symbols}\\
$\pSI_t$: System imbalance price [€/MWh]\\
$\pMDP_t$: Downregulaton price [€/MWh]\\
$\pMIP_t$: Upregulation price [€/MWh]\\
$s_t$: System imbalance volume [MW]\\
$\fSI_t$: pdf of the system imbalance price\\
$\fMDP_t$: pdf of the downregulation price\\
$\fMIP_t$: pdf of the upregulation price\\
$\pi_t$: Mixture weight model \\
$n_q$: Number of quantile models in the regulation price model \\
$V_\Rtwo$: Discretized range of aFRR reserve volumes \\
$V_\Rthree$: Discretized range of mFRR reserve volumes \\
$\beta$: Market reactivity parameter\\
$u$: Trade position [MW] \\
$q_t$: Intraday price [€/MW] \\
$K_\MDP$: Downregulation price sensitivity [€/MW]\\
$K_\MIP$: Upregulation price sensitivity [€/MW]\\
$\alpha$: Risk tuning parameter\\
$u^*_{\rho_\alpha, t}$: Optimal trade position w.r.t. risk measure $\rho_\alpha$\\
$N$: Window size of historical trades

\section{Introduction}
The rising share of renewable energy sources is driving rapid evolution in electricity markets. Their inherent volatility and unpredictability, largely due to weather, make it more challenging for transmission system operators (TSOs) to balance consumption with production \cite{GOODARZI2019110827}. Rather than solely relying on the activation of traditional energy reserves, TSOs are increasingly engaging \textit{balance responsible parties} (BRPs) to actively aid in balancing the grid.
To facilitate this effort, TSOs utilize imbalance settlement schemes where BRPs are rewarded for maintaining the balance and penalized for creating an imbalance. In this context, accurately predicting future system imbalances becomes a crucial decision-making tool. 
By anticipating future energy shortages or surpluses in the local balancing market, electricity may be strategically purchased on the intraday market and used to balance the system for profit. This paper delineates an automated trading strategy based on this principle. The strategy compromises two main steps: (i) modelling future system imbalance prices, and (ii) converting these predictions into profitable positions on the intraday electricity market. The intraday market is used to purchase electricity shortly before it closes; therefore, we do not focus on adjusting our position throughout the day or modelling the intraday price itself. This section provides a brief introduction into electricity markets, discusses related work, and outlines the structure and key contributions of this paper.

\subsection{The Electricity Market}\label{sec:energy-market}
\begin{figure}
  \centering
  \includegraphics[width=88mm]{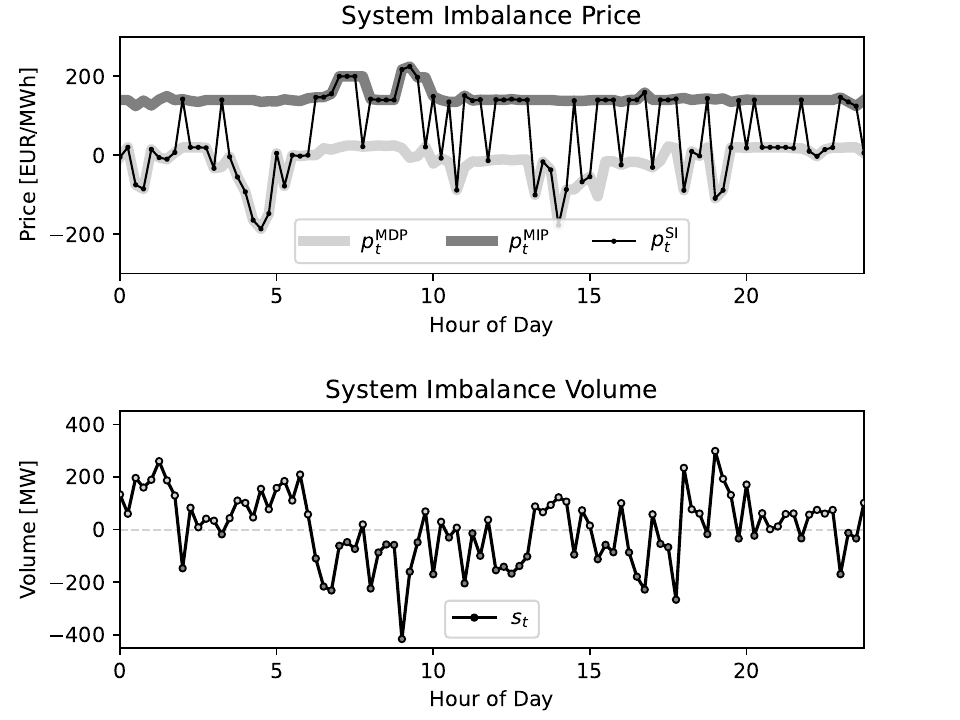}
  \caption{System imbalance price and volume in Belgium on May 26 2023. Depending on the sign of the system imbalance volume $s_t$, the system imbalance price $\pSI_t$ continuously alternates between the upregulation and downregulation price. This data has been obtained from the Belgian TSO Elia.
}
  \label{fig:si-price}
\end{figure}

Modern electricity markets consist of several components. The largest volumes are traded on the day-ahead market, where transactions occur one day before delivery. Nowadays, this market is typically complemented by an intraday market which enables balance responsible parties (BRPs) to adjust their positions in the hours leading up to delivery. On these intraday markets, electricity may be traded through auctions (pay-as-cleared) or via continuous trading (pay-as-bid) \cite{epexspot_power_market_basics}. 
Despite intraday trading efforts, perfectly balancing the actual production and consumption is impossible. The combined imbalance of all BRPs connected to the local grid is referred to as the \textit{system imbalance volume} and to compensate for this imbalance, TSOs activate energy reserves.
The costs associated with the activation of these reserves are reflected in a \textit{system imbalance price}, which is a price used to settle energy imbalances among BRPs. Those who help balance the grid are rewarded and those who create the imbalance are penalized. 
The TSO is responsible for determining this system imbalance price and volume, as well as managing the settlement of energy imbalances. \\
Various methodologies for calculating this system imbalance price exist.
The European Union Agency for the Cooperation of Energy Regulators (ACER) advocates for single-price imbalance settlement schemes where the price for having a positive or negative energy imbalance is the same \cite{acer2024monitoring}. This incentivizes market participants to actively aid in balancing the grid. In contrast, dual-price settlement schemes impose additional costs for imbalances making the rewards for contributing smaller than the penalties. This encourages participants to solely focus on minimizing their own imbalance rather than supporting overall grid stability. \\ 
In this paper, we specifically focus on the Belgian electricity market, which features a single-price imbalance settlement scheme and is characterized by a high share of renewable energy sources. In the Belgian electricity market,
a single system imbalance price $\pSI_t$ is settled every quarter hour (denoted by time index $t$) based on the average system imbalance volume $s_t$ during that interval:
\begin{equation}\label{eq:si-price}
  \pSI_t(s_t) = \left\{ \begin{array}{ll}
    \pMDP_t & \textrm{if} \quad s_t \ge 0, \\
    \pMIP_t & \textrm{if} \quad s_t < 0.
  \end{array} \right.
\end{equation}
Here, $\pMDP_t$ represents the downregulation price or the price associated with the activated downward reserves in case of an energy surplus ($s_t \ge 0$) and $\pMIP_t$ the upregulation price or the price associated with the activated upward reserves when there's an energy shortage ($s_t < 0$) \cite{baetensImbalancePricingMethodology2020}. Typically, the downregulation price, $\pMDP_t$, is low or even negative, as it reflects the need to eliminate an energy surplus. In contrast, the upregulation price, $\pMIP_t$, is high because it indicates the need to boost production. As visualized in Figure \ref{fig:si-price}, this formulation leads to a system imbalance price that continuously alternates between the two regulation prices.
The Belgian electricity market is linked to the European intraday exchange where energy is continuously and freely traded on the interconnectors with neighbouring countries, as long as sufficient border capacity is available. The cross-border intraday market closes exactly one hour before delivery. 

\subsection{Related Work}
\subsubsection{Modelling future system imbalance prices}\label{sec:related-work-model}
The main difficulty in modelling the Belgian system imbalance price resides in its inherent alternating behaviour. 
The prices exhibit a bimodal distribution, which traditional predictive models often fail to capture. 
In the literature, two methodologies to predict system imbalance prices can be distinguished:
(i) those that model the balancing state implicitly, and (ii) those that model it explicitly.
Implicitly modelling the balancing state involves employing a single model to predict balancing prices, and in doing so, assumes that the typical alternating behaviour is automatically discovered by the model \cite{narajewskiProbabilisticForecastingGerman2022,limaBayesianPredictiveDistributions2023,makriShortTermNet2021}.
Conversely, explicitly modelling the balancing state entails the use of multiple models to predict balancing prices. Typically, one model predicts the probability of arriving in a future balancing state, while separate models compute the prices for each balancing state, thereby reducing the overall model complexity. \\
A popular approach to explicitly model the balancing state are regime-switching Markov models \cite{bunnAnalysisFundamentalPredictability2021,klaeboeBenchmarkingTimeSeries2015,olssonModelingRealTimeBalancing2008,dumasProbabilisticForecastingImbalance2019}. Future balancing states are modelled as a Markov chain, and a distinct price is forecasted for each balancing state. Despite its popularity, a straightforward implementation of regime-switching Markov models in this study reveals two notable issues: (i) static state transition probability matrices are used, independent of other external inputs which might be relevant, and (ii) this method is not suited for longer prediction horizons since the state probabilities converge to a stationary distribution.\\
Instead of using Markov chains, Koch proposes a logistic regression model directly predicting the probability of a future balancing state \cite{kochIntradayImbalanceOptimization2022}. This alternative approach mitigates the need to model all intermediate balancing states, rendering it particularly useful for longer prediction horizons. Unlike the static state transition probability matrices, the logistic regression model can easily incorporate other relevant external inputs.

\subsubsection{Risk-Averse Trading Strategy}
Since the introduction of intraday trading, studies have demonstrated that trading contributes to the balance of the grid \cite{KOCH2019109275,POPLAVSKAYA2020115130,LAGO2021110467}. There are several ways to trade on the intraday market; however, in this study, we solely operate as a price-taker on the continuous intraday market just before the cross-border market closes. 
Given that the distribution of future prices was exactly known, one could 
simply select a trading strategy that maximizes the expected future return. 
In practice, however, these distributions are estimated with models that 
were fit to a finite amount of training data, and are therefore subject 
to some amount of error. 
When applied naively, this typically leads to larger trading losses than predicted.
A popular approach for taking this model distributional uncertainty -- also known 
as ambiguity -- into account, is to replace the expectation by a \emph{coherent risk measure}
\cite{shapiroLecturesStochasticProgramming2009}.
Based on the choice of risk measure, larger realizations of the cost will be 
penalized more, resulting in a more risk-sensitive behavior. Furthermore, 
coherent risk measures can be interpreted as robust counterparts to the expectation. Instead of minimizing the cost over the predicted distribution, the addition of the risk measure can be viewed as a minimization over a specific set of distributions, known as the \emph{ambiguity set}
\cite[Thm. 6.4]{shapiroLecturesStochasticProgramming2009}. 
As a result, this approach robustifies the trading strategy against misestimations 
of the predicted cost distribution. \\ 
Risk-averse optimization is a decision making tool that has been used in power systems before \cite{bruninxInteractionAggregatorsElectricity2020,rashidizadeh-kermaniRegretBasedStochasticBiLevel2020,ghavidelRiskConstrainedBiddingStrategy2020,toubeauMediumTermMultimarketOptimization2018}.
Typically, the tuning parameters are selected in an offline fashion. 
Bottiau et al. pointed out that using a machine learning model to dynamically 
adjust the tuning parameters of the risk measure resulted in better performance when applied to very-short-term energy markets \cite{bottieauAutomaticRiskAdjustment2021}. 
With online learning, the machine learning model is able to adaptively learn the model mismatch. 
In recent years, 
model-free alternatives relying entirely on machine learning methods have gained 
significant popularity, as they relieve the need for precise modeling 
of often complex market behaviors.
In particular, deep reinforcement learning has emerged as a
powerful technique for optimizing bidding strategies and asset management in the electricity market \cite{HUA10330727}. 
The main disadvantage of these techniques, however, is that their behavior 
remains largely unpredictable and difficult to interpret.
Instead, we propose a hybrid model, where certain aspects
of the market are directly encoded into the model, while others are learned in a data-driven manner, 
leading to a more efficient use of the available data.

\subsection{Contribution and Structure}
In this paper, an automated intraday electricity trading strategy for single-price balancing markets is proposed.
The contributions of this paper are summarized as:
\begin{itemize}
  \item A novel probabilistic mixture model to predict future price distributions in single-price balancing markets (cf. \S\ref{sec:models}). 
  \item A risk-averse trading strategy to reduce the number of trades in uncertain scenarios  (cf. \S\ref{sec:trading_strat}). The tuning parameters are optimized in an online fashion based on a window of recently observed data, and the strategy explicitly accounts for the influence of the position on the imbalance price.
\end{itemize}
In \S\ref{sec:case-study}, the model is trained on data from the Belgian electricity market. After benchmarking against several models, the trading strategy is validated with a simulation using real-time data. Finally, in \S\ref{sec:discussion}, the results are discussed.

\subsection*{Notation}
Let $\R$ ($\R_{+}$) denote the (nonnegative) reals.
We denote the $n$-dimensional probability simplex as 
$\simplex_{n} = \{ p \in \R^{n}_{+} \mid \sum_{i=1}^{n} p_i = 1\}$.
We introduce the shorthand $[n] = \{1, \dots, n\}$ to denote the $n$ smallest positive natural numbers.
We denote the positive part of $x$ by $[x]_+ = \max\{x, 0\}$. 

\section{Mixture Imbalance Price Model}\label{sec:models}
\begin{figure}
  \centering
  \includegraphics[width=88mm]{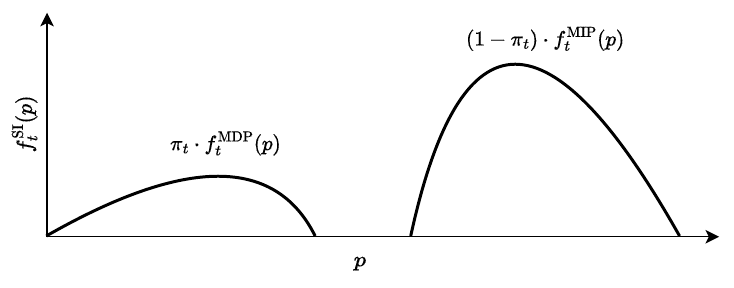}
  \caption{A sketch of the pdf of the future system imbalance prices $\fSI_t$. Using a mixture model, we relieve the model from having to predict this bimodal distribution directly. Instead, we predict the regulation price distributions $\fMDP_t$ and $\fMIP_t$ separately, and combine these together with the mixture weight $\pi_t$.}
  \label{fig:two-step-model}
\end{figure}

To properly quantify the trading risks and profits, we are interested in predicting the \ac{pdf} of the future system imbalance price. As mentioned in \S\ref{sec:energy-market}, this price continuously alternates between the two regulation prices, causing the future price to be bimodally distributed. 
We explicitly take this bimodality into account by modelling the 
\ac{pdf} of the imbalance price as a mixture distribution \cite{mclachlanFiniteMixtureModels2019}.
In this context, the \ac{pdf} $f: \Xi \to \R_+$ is said to be a mixture of a finite set of component densities $f_i(\,\cdot\,; \ve{\theta}_i)$ with parameters $\boldsymbol{\theta}_i$, $i \in [d]$. A mixture distribution can be written in the form
\begin{equation} \label{eq:mixture-density}
    f(\xi) = \sum_{i=1}^{d} \gamma_i f_i(\xi; \ve{\theta}_i)
\end{equation}
where $\ve{\gamma} \in \simplex_{\nmix}$ is a vector of nonnegative mixture weights that sum to one.
A major advantage of mixture distributions is their ability to represent complex distributions as a combination of simpler component distributions. This approach is particularly effective for modelling multimodal behavior, 
such as the alternating dynamics of the imbalance price.
A popular choice for the component densities is the normal distribution $\mathcal{N}(\mu_i, \sigma_i)$ with $\ve{\theta}_i = (\mu_i, \sigma_i)$.
Mixture models are often used when samples are observed from a mixture of distributions without knowing from which component distribution they originate. 
The components are considered \emph{latent variables} and,  
in that case, specialized algorithms, such as Expectation-Maximization are used 
to jointly estimate the component weights $\gamma_i$ and the parameters $\ve{\theta}_i$ 
of the distributions, $i \in [d]$ \cite{mclachlanFiniteMixtureModels2019}.
Fortunately, in our application, this difficulty does not arise, 
as it will be straightforward to infer the active component for each 
observed data point.\\
Indeed, in the context of this work, mixture distributions are employed to incorporate prior knowledge of the system imbalance price settlement scheme into the model. As defined in Eq. \eqref{eq:si-price}, the system imbalance price $\pSI_t$ is determined by the sign of the system imbalance volume $s_t$. Specifically, when the imbalance volume is positive, the system imbalance price equals the downregulation price $\pMDP_t$, whereas if it is negative, it matches the upregulation price $\pMIP_t$. To simplify the modeling task, we model the system imbalance price as a mixture of the two regulation price distributions. Here, the sign of the system imbalance volume acts as an exogenous variable indicating which density component is active. 
Using Bayes' theorem, the probability density function of the future system imbalance price $\fSI_t(p)$ at time $t$ is expressed as the following mixture distribution:
\begin{equation}\label{eq:si_distribution}
  \fSI_t(p) = \pi_t \cdot \fMDP_t(p) + (1 - \pi_t) \cdot \fMIP_t(p)
\end{equation}
with
\begin{equation}\label{eq:si_conditions}
  \begin{aligned}
    \pi_t &= \prob(s_t > 0)\\
    \fMDP_t(p) &= \fSI_t(p \mid s_t > 0)\\ 
    \fMIP_t(p) &= \fSI_t(p \mid s_t \le 0)
  \end{aligned}
\end{equation}
where $s_t \in S_t = \R$ is a random variable representing the future system imbalance volume,
and $\fSI_t(\,\cdot \mid E)$ denotes the conditional \ac{pdf} of the price $p_t$ at time $t$, given event $E$.
Thus, $\pi_t$ is the probability to obtain a positive system imbalance volume, $\fMDP_t$ is the conditional probability density function of the system imbalance price given a positive system imbalance volume, and $\fMIP_t$ the conditional probability density function of the system imbalance price given a negative system imbalance volume. To model future system imbalance price distributions, we may choose to model these components separately to reduce the overall complexity of the problem. 
A visual representation of this mixture distribution is given in Fig. \ref{fig:two-step-model}.
It is important to note that the mixture weight and distributions are time-dependent (e.g. higher probability on energy shortage and more expensive prices during peak hours), 
yet, only one system imbalance price $p_t$ is observed at each timestep $t$. 
That is, we only observe a single sample from the distribution $\fSI_t(p)$, 
making it practically impossible to estimate the probability density function directly.
To address this issue,
we assume that the time dependence is introduced through measurable inputs 
$\ve{x}_t \in \R^{\nx}$. 
With this assumption, we can parametrize the mixture distribution as time-invariant functions with parameters $\ve{\theta}_\pi$, $\thetaMDP$ and $\thetaMIP$:
\begin{align*}
    \pi_t &= \pi(\ve{x}_t, \ve{\theta}_{\pi}) \\
    \fMDP_t(p) &= \fMDP(p; \ve{x}_t, \thetaMDP) \\
    \fMIP_t(p) &= \fMIP(p; \ve{x}_t, \thetaMIP).
\end{align*}
These time-invariant models can be estimated with statistical methods.
In the upcoming sections, we elaborate on the choice of these models. The mixture weight is predicted with a logistic regression model, while the regulation price distributions are predicted with non-parametric probabilistic models.

\subsection{Mixture Weight Model}\label{sec:prob}
The mixture weight in Eq. \eqref{eq:si_distribution}, denoted by $\pi_t = \prob(s_t > 0)$, represents the probability of having a positive system imbalance at timestep $t$.
Various approaches have been proposed for the problem of estimating the probability of such a 
binary event \cite{niculescu-mizilPredictingGoodProbabilities}. 
One of these approaches is logistic regression, an elementary method used to 
predict the probability of an event.
Logistic regression has already been proven successful in forecasting future balancing states in the German electricity market, motivating its application in this study \cite{kochIntradayImbalanceOptimization2022}. 
Formally, the logistic regression model is a sigmoid function applied to a linear model:
\begin{equation}\label{eq:log_reg}
  \begin{aligned}
      \pi_t = \pi(\ve{x}_t, \ve{\theta}_\pi) = \sigma(\ve{w}^\top\ve{x}_t + w_0) = \frac{\exp(\ve{w}^\top \ve{x}_t + w_0)}{1+\exp(\ve{w}^\top \ve{x}_t + w_0)}
  \end{aligned}
\end{equation}
where the weights $\ve{\theta}_{\pi} = (w_0, \ve{w}) \in \R^{\nx+1}$
are optimized with maximum-likelihood estimation. The simplicity of this model renders it robust to overfitting.
Theoretically, this model also requires the log odds of an event to be a linear combination of the input variables \cite{cucchiaraAppliedLogisticRegression2012}.
Despite this condition usually not being satisfied, logistic regression still yields meaningful results in practice. 

\subsection{Regulation Price Model} \label{sec:price-model} 
The regulation price models should output the distribution of the regulation prices.
Models for probability distributions may either be parametric or non-parametric.
Parametric probabilistic models are trained to output the parameters of a predefined family of distributions (e.g., the mean and variance of a normal distribution) and are particularly useful if the distribution type is known beforehand.
When this is not the case, non-parametric probabilistic models are typically preferable.
These models predict multiple quantiles of an output variable to build a discrete
approximation of the \ac{cdf}, which fully characterizes the distribution of a scalar random variable.
Since (i) there is no obvious choice for a parametric family of distributions 
that describes the complex nature of regulation price,
and (ii) the random variable is scalar, so a fine discretization of its support is computationally feasible,  
non-parametric models are used in this paper.
Before introducing the price model, we 
briefly explain the energy reserves, a relevant input signal to determine the regulation price distribution.
\subsubsection{Energy Reserves}
To compensate for energy shortages or surpluses, TSOs activate reserves. For example, in Belgium, two types of reserves can be activated: automatic reserves (aFRR) and manual reserves (mFRR). The aFRR reserves are activated almost instantly, but are limited in volume. The mFRR reserves, while slower to activate, are larger in volume. Depending on the size of the imbalance, a combination of these reserves may be activated. 
In Belgium, the costs of activating a certain amount of these reserves in the future are known beforehand and can be used as a mapping from reserve volume to price.
We propose to predict the regulation price by internally modelling the activated volumes of each reserve type
based on input features $\ve{z}_t$.
Using the aforementioned price mapping, the corresponding reserve price is then calculated.  

\subsubsection{Regulation Price Model} \label{subsec:price-model}
\begin{figure}
  \centering
  \includegraphics[width=88mm]{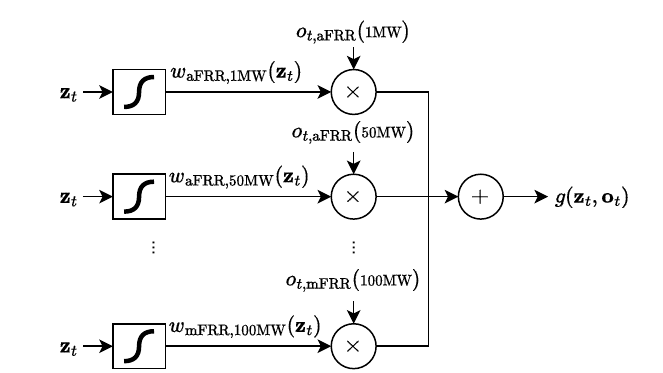}
  \caption{A visual representation of the price model. 
      The softmax operation ensures that $\ve{w}(\ve{z}_t)$ represents a valid probability 
      distribution, with nonnegative elements that sum to one. 
  This probability vector is then multiplied with the order book prices $\ve{o}_t$ to obtain a prediction of the regulation price $g(\ve{z}_t, \ve{o}_t)$.}
  \label{fig:price_model}
\end{figure}

For each reserve type $r \in R = \{\mathrm{aFRR}, \mathrm{mFRR}\}$, 
we discretize the range of reserve volumes $v \in V_r = \big\{v_{r}^{(1)}, \dots, v_{r}^{(n_r)}\big\}$
and use the available data to estimate a discrete probability distribution
\begin{equation} \label{eq:prob-vec}
    \ve{w}(\ve{z}_t, \ve{\theta}) = \big( w_{r, v}(\ve{z}_t, \ve{\theta}) \big)_{v \in V_r, r \in R} \in \simplex_{n_{\Rtwo} + n_{\Rthree}}, 
\end{equation}
as a function of the input features $\ve{z}_t$ and model parameters $\ve{\theta}$. Here,  
$w_{r, v}(\ve{z}_t, \ve{\theta})$ represents the probability that
the imbalance price is equal to the price of
a given volume $v \in V_r$ of reserve $r$, 
given the input $\ve{z}_t$.
Using the price mappings $o_{t, r}: V_r \to \R$, $r \in R$ at time $t$,
we then predict the expected price for the activation of the reserves as
\begin{equation}\label{eq:price_model}
    \hat{g}_{\ve{\theta}}(\ve{z}_t, \ve{o}_t) = \langle \ve{w}(\ve{z}_t, \ve{\theta}), \ve{o}_{t} \rangle, 
\end{equation}
where $\ve{o}_t$ is a vector containing the (known) prices for activating a volume of $v \in V_r$ of reserve $r$
at time $t$: 
$$\ve{o}_{t} = \big(o_{t, r}(v)\big)_{v \in V_r, r \in R} \in \R^{n_{\Rtwo} + n_{\Rthree}}$$
The probability vector $\ve{w}_t$ in Eq. \eqref{eq:prob-vec} is modelled with a linear model followed by a softmax activation function. This approach, also known as multinomial logistic regression, is widely used for predicting the probabilities of multiple outcomes \cite{bhningMultinomialLogisticRegression1992}.
The structure of the model is visually depicted in Fig. \ref{fig:price_model}. 

We use the model structure from Eq. \eqref{eq:price_model}
to obtain an estimate of the \ac{cdf} of both the upregulation and the downregulation price.
To this end, we first discretize the interval $[0,1]$ into $n_q$ equally spaced 
values $\{\tau_{1}, \dots, \tau_{n_q}\}$. 
For both regimes $m \in \{\mathrm{MDP}, \mathrm{MIP} \}$, and 
for each $i = 1, \dots, n_q$, our goal is to 
find the model parameters $\ve{\theta}_{m,i}$,
such that \(g_{i}^{m}(\ve{z}_t, \ve{o}_t) = \hat{g}_{\ve{\theta}_{m,i}}(\ve{z}_t, \ve{o}_t)\)
approximates the $\tau_{i}$'th quantile of the up- or downregulation 
price (depending or $m$).
We achieve this using a standard method known as quantile regression, which 
for a given quantile level $\tau \in [0, 1]$ is
equivalent to a classical regression problem from $\ve{z_t}$ to $p_t$
with a particular loss function, known as the \textit{quantile loss}
or \textit{pinball loss} \cite{koenkerQuantileRegression2001}, i.e.,
\begin{equation}\label{eq:quantile}
    L_\tau (e) = 
    \begin{cases}
        \tau e & \textrm{if } e \ge 0\\
        (\tau - 1) e & \textrm{if } e < 0.
    \end{cases}
\end{equation}
By minimizing this loss function, the optimal parameters are found.
Occasionally, the estimated 
\acp{cdf} may not be monotone increasing with $\tau$. In such cases, we enforce this constraint by reordering the quantiles.
Since the quantiles are evenly spaced, the corresponding probability distribution can be represented
by a sum of discrete point masses, each with a probability of $1/n_q$:
\begin{equation}\label{eq:probability-density}
  \begin{aligned}
    \fMDP_t(p) &= \frac{1}{n_q} \sum_{i=1}^{n_q} \delta\left(p - g_i^\MDP(\ve{z}_t, \ve{o}_t) \right)\\
    \fMIP_t(p) &= \frac{1}{n_q} \sum_{i=1}^{n_q} \delta\left(p - g_i^\MIP(\ve{z}_t, \ve{o}_t) \right)
  \end{aligned}
\end{equation}
Here, the function $\delta$ represents the Dirac distribution,
which places unit probability mass at the origin and is zero elsewhere \cite[Ex. 4.7]{schilling_MeasuresIntegralsMartingales_2005}. 

\section{Trading Strategy}\label{sec:trading_strat}
Electricity markets, much like stock markets, enable traders to take both long and short positions. In this paper, a long position involves purchasing energy on the intraday market and settling the resulting imbalance on the balancing market, yielding a profit if the expected difference between the imbalance price and the intraday price is positive. Conversely, a short position is profitable when this price difference is negative. In this study, we consistently act as price-takers on the continuous intraday market. By examining the order books, we can identify the prices at which we can instantly buy or sell a specific amount of energy.
In this section, we begin by formally defining how trade positions impact the imbalance price. Thereafter, we introduce the risk-averse trading strategy and present a methodology for adaptively tuning the risk parameters.

\subsection{Impact on the Balancing Market}\label{sec:market-impact}
We assume that intraday trading positions are fully settled on the balancing market. That is, taking an intraday long position results in additional energy being produced or imported into Belgium, which naturally influences both the system imbalance volume and price.
In this section, we present a simple model to take this effect into account.

\begin{figure}
  \centering
  \includegraphics[width=88mm]{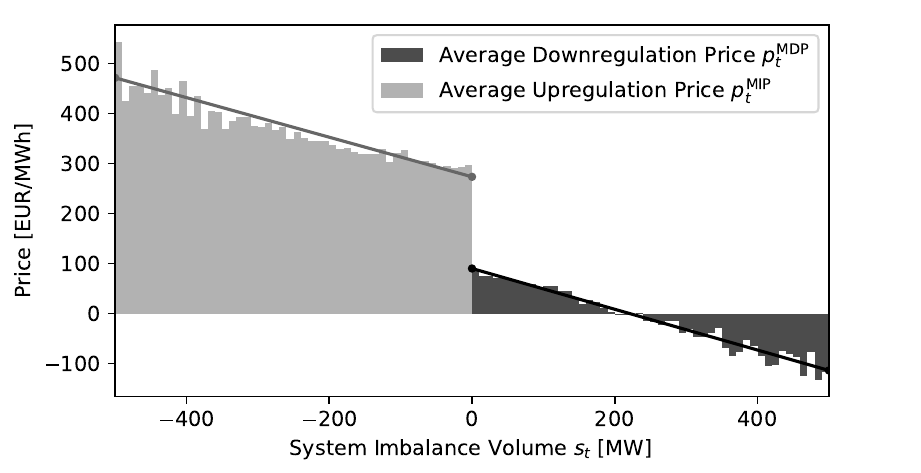}
  \caption{The average regulation price for different system imbalance volumes on the Belgian balancing market between May 2021 and May 2023. The sensitivity of the regulation prices to the system imbalance volume is approximated by fitting two linear slopes. This seems like a reasonable approximation.}
  \label{fig:regulation-price-sensitivity}
\end{figure}

For a long position of size $u$, we assume that the system imbalance volume will increase with a volume $\beta u$, where $\beta \in [0, 1]$ is a tuning parameter 
that models the market reactivity. In an open-loop system, where BRPs do not react to the actions of others, $\beta$ would be equal to one. However, in reality, the activations of one party would lead to fewer activations in the same direction by others ($\beta < 1$).
Accurately estimating $\beta$ is highly challenging, as it requires placing numerous trades as a trader and detecting changes in the behavior of the system imbalance volume. 
For simplicity, we therefore consider it a tuning parameter of the method.
To ensure meaningful simulations in our case study, we consider the most pessimistic scenario where our traded volume has the strongest negative impact on the market price by setting $\beta=1$. Indeed,
replacing the imbalance volume $s_t$ 
by $s_t + \beta u$, the system imbalance price according to Eq. \eqref{eq:si-price} becomes
\begin{equation}
  \pSI_t(s_t + \beta u) = \left\{ \begin{array}{ll}
    \pMDP_t(s_t + \beta u) & \textrm{if} \quad s_t + \beta u \ge 0 \\
    \pMIP_t(s_t + \beta u) & \textrm{if} \quad s_t + \beta u < 0
  \end{array} \right.
\end{equation}
If $u$ is sufficiently large, it will not only impact the imbalance volumes, but 
the regulation prices as well.
Based on empirical observations (cf. Fig.~\ref{fig:regulation-price-sensitivity}),
    we propose to model the impact of our position on the regulation prices using a piecewise-linear approximation
\begin{equation}\label{eq:modified-si-price}
    \pSI_t(s_t + \beta u) = 
    \begin{cases}
    \pMDP_t(s_t) - K_\mathrm{MDP} \beta u  & \textrm{if} \quad s_t + \beta u \ge 0 \\
    \pMIP_t(s_t) - K_\mathrm{MIP} \beta u & \textrm{if} \quad s_t + \beta u < 0
    \end{cases}.
\end{equation}
Here, the constants $K_\mathrm{MDP}$ and $K_\mathrm{MIP}$ represent the sensitivity of the regulation price to the system imbalance volume. The larger these constants, the lower the system imbalance price will be and therefore also the profit of the trading strategy. 
We estimate these sensitivities by fitting the slope between the system imbalance volume and regulation price with a least squares regression:
\begin{equation}
  \begin{aligned}
  K_\MDP &= \frac{\sum_{t \in T_\MDP} (s_t - \bar{s}_t^\MDP) (\pMDP_t - \bar{p}_t^\MDP)}{\sum_{t \in T_\MDP} (s_t - \bar{s}_t^\MDP)^2} \\
  K_\MIP &= \frac{\sum_{t \in T_\MIP} (s_t - \bar{s}_t^\MIP) (\pMIP_t - \bar{p}_t^\MIP)}{\sum_{t \in T_\MIP} (s_t - \bar{s}_t^\MIP)^2}
  \end{aligned}
\end{equation}
where $T_\MDP$ and $T_\MIP$ are the sets of timesteps where the system imbalance volume is positive and negative, respectively. The terms $\bar{s}_t^x$ and $\bar{p}_t^x$ are the mean system imbalance volume and price over $T_x$. Fig. \ref{fig:regulation-price-sensitivity} visualizes the price sensitivity of the regulation prices to the system imbalance volume on the training set of the case study, where we obtain $K_\mathrm{MDP} = 0.40$ €/MW and $K_\mathrm{MIP} = 0.41$ €/MW. This implies that, on average, an increase of \SI{1}{\mega\watt} in the system imbalance volume leads to a decrease of €0.41 in the upregulation price. \\
The modified imbalance price from Eq. \eqref{eq:modified-si-price} is used to build a position-aware model $\fSI_{t,u}$ that predicts the imbalance price distribution after placing a trade of size $u$.
The modified mixture model now has to predict the probability that the system imbalance exceeds $-\beta u$:
\begin{equation}
  \pi_{t}(u) = \mathbb{P}(s_t + \beta u > 0) = \mathbb{P}(s_t > -\beta u)
\end{equation}
In this paper, the input $u$ is integrated as an additional feature in the logistic regression model from Eq. \eqref{eq:log_reg}. To achieve this, the training set is augmented with artificial trade positions $u$ which are sampled from a uniform distribution over the interval $U = \left[0, u_{\max} \right]$. The model is then trained to predict the probability that the system imbalance volume exceeds $-\beta u$.
For the regulation price models, 
we simply train the quantile models using the observed regulation prices $\pMDP_t(s_t)$, $\pMIP_t(s_t)$ and then subtract $K \beta u$ from each predicted quantile. The following expression represents the position-adjusted pdf of the system imbalance price:

\begin{equation}\label{eq:modified-si-distribution}
  \fSI_{t,u}(p) = \pi_{t}(u) \cdot \fMDP_{t,u}(p) + (1 - \pi_{t}(u)) \cdot \fMIP_{t,u}(p)
\end{equation}
with
\begin{equation}
  \begin{aligned}
    \pi_{t}(u) &= \prob(s_t > - \beta u) = \pi\big((\ve{x}_t, u), \ve{\theta}_\pi\big) \\
    \fMDP_{t,u}(p) &= \fMDP_t(p + K_\MDP \beta u) \\ 
    \fMIP_{t,u}(p) &= \fMIP_t(p + K_\MIP \beta u)
  \end{aligned}
\end{equation}
This formulation may also be used for a short position by substituting $u$ with a negative value and training a similar model.

\subsection{Risk-Averse Trading Strategy}\label{sec:risk-averse-strat}
The ultimate goal of an energy trader is to find the trade position on the electricity market that maximizes the expected profit,
or equivalently,
the position that minimizes the expected loss.
To ease the notation of the risk measures, we continue with minimization problems.
The loss $\Z{p,t}{u}$ for a long strategy with position size $u \in U \subset \R_+$ and delivery time $t$ may be expressed as:
\begin{equation*}
  \Z{p,t}{u} = \big(q_{t}(u) - p \big)u,\quad p \in P_t
\end{equation*} where the random variable $p \in P_t$ denotes the future system imbalance price and $q_{t}(u)$
denotes the current (known) intraday price\footnote{The intraday price may arise from filling multiple orders at different price levels. In that case, the intraday energy cost $q_{t}(u) u$ is a piecewise-linear and monotone-increasing function of the position size $u$, as it is derived from the integration of piecewise-constant, increasing prices.} to take a position of size $u$. 
The optimal position $u_t^*$ that minimizes the expected loss then equals
\begin{equation}  \label{eq:true-problem}
    u_t^* = \argmin_{u \in U} \mathbb{E}_{p}[\Z{p,t}{u}], 
\end{equation}
where the expectation is taken with respect to the random variable $p$.
While $q_{t}(u)$ is known exactly from the order book, the true distribution of the future imbalance price $p$ is not.
Using the estimated \ac{pdf} $\fSI_{t,u}$, from Eq. \eqref{eq:modified-si-distribution}, 
we may replace the expectation in Eq. \eqref{eq:true-problem} by an approximation
\begin{equation*}
    u_t^* \approx \argmin_{u \in U} \mathbb{E}_{p \sim \fSI_{t,u}}[\Z{p,t}{u}]. 
\end{equation*}
However, with this methodology, the strategy ignores the fact that the price distribution was replaced by a prediction.
This may lead to unexpected returns of the strategy when the
predicted prices deviate a lot from the true distribution. 
Instead, we introduce risk-awareness by replacing the expectation with a coherent risk measure $\rho_\alpha$:
\begin{equation}\label{eq:risk_averse}
  \begin{aligned}
      u^*_{\rho_\alpha,t} &= \argmin_{u \in U} \rho_{\alpha,\,p \sim \fSI_{t,u}}[\Z{p,t}{u}].
  \end{aligned}
\end{equation}
The tuning parameter $\alpha$ controls the level of risk aversion, allowing the trader to balance the desire for higher profits with the need to mitigate risk.
An interesting property of coherent risk measures is that they can be viewed as a worst-case expected value over a specific set of distributions, 
known as the \emph{ambiguity set} \cite[Thm. 6.4]{shapiroLecturesStochasticProgramming2009}.
That is, instead of minimizing the expected loss over a single price distribution 
$\fSI_t$, we consider a set of similar distributions $\amb_{t,u}(\alpha)$ to account for potential inaccuracies in the cost:
\begin{equation*}
    u^*_{\rho_\alpha,t} = \argmin_{u \in U} \normmax_{f \in \amb_{t,u}(\alpha)} \mathbb{E}_{p \sim f} [\Z{p,t}{u}]
\end{equation*}
By taking the worst-case realization of the expected loss in this ambiguity set, the trading strategy is made robust against all realizations within this set.
The size of this ambiguity set varies with the choice of $\alpha$. 
A larger ambiguity set leads to a more conservative trading strategy that is less sensitive to potential modelling errors, 
while a smaller ambiguity set leads to a more aggressive trading strategy being more vulnerable to these errors.
In this paper, we discuss two commonly used coherent risk measures, namely
conditional value at risk and entropic value at risk \cite{rockafellarOptimizationConditionalValueatrisk2000,ahmadi-javidEntropicValueatRiskNew2012}.\\
The conditional value-at-risk ($\CVAR$), 
also known as the expected shortfall, equals the expected loss in the worst $\alpha$ cases. 
$\CVAR$ may be expressed as a convex optimization problem:
\begin{equation}\label{eq:cvar}
  \CVAR_{1-\alpha}(Z) =  
\begin{cases}
    \essmax Z & \text{ if } \alpha = 0\\ 
    \inf_{s} \left\{ s + \frac{1}{\alpha} \mathbb{E} \left[ Z - s\right]_+ \right\} & \text{ otherwise}
\end{cases}
\end{equation}
where $\alpha \in [0,1]$. For $\CVAR$, this problem can be solved analytically to obtain the average loss in the worst $\alpha$ cases.
By considering only the worst cases, a higher weight is given to the losses making the trading strategy more risk sensitive. This risk measure converges to the expectation (risk-seeking) when $\alpha$ approaches 1. Conversely, when $\alpha$ approaches 0, this risk measure converges to the maximum value of $Z$ or the largest loss (risk-averse). \\
The entropic value-at-risk ($\EVAR$) is defined as the following convex optimization problem: 
\begin{equation} \label{eq:evar}
  \EVAR_{1 - \alpha} (Z) = \begin{cases}
    \essmax Z & \text{ if } \alpha = 0\\ 
    \inf\limits_{s > 0} \left\{ 
    \frac{1}{s} \ln \left( \frac{\mathbb{E}[e^{sZ}]}{\alpha} \right)
   \right\} & \text{ otherwise}
  \end{cases}
\end{equation}
Similar to $\CVAR$, $\EVAR$ also interpolates between the expectation 
and the maximum value as $\alpha$ decreases from 1 to 0. 
An interesting property of coherent risk measures 
is their positive homogeneity and translation invariance 
\cite{ahmadi-javidEntropicValueatRiskNew2012}, 
i.e., 
\begin{align*}
  \rho_\alpha[a Z + b] &= a\,\rho_\alpha[Z] + b, \quad \forall a \in \R_+,\forall b \in \R.
\end{align*}
Using this identity, Eq. \eqref{eq:risk_averse} may be reformulated as
\begin{equation}\label{eq:solution}
  \begin{aligned}
    u^*_{\rho_\alpha,t} &= \argmin_{u \in U} \rho_{\alpha,\,p \sim \fSI_{t,u}} [\Z{p,t}{u}] \\
          &= \argmin_{u \in U} \rho_{\alpha,\,p \sim \fSI_{t,u}}  [(q_{t}(u)- p) u] \\
          &= \argmin_{u \in U} \underbrace{\left( q_{t}(u) + \rho_{\alpha,\,p \sim \fSI_{t,u}} [-p] \right) u}_{= \varphi(u)}.
  \end{aligned}
\end{equation}
The reference distribution $\fSI_{t,u}$ depends on the decision variable 
$u$, making this problem an instance of stochastic optimization with decision-dependent distributions.
This class of problems is generally difficult to solve even for risk measures as simple as the expectation \cite{drusvyatskiyStochasticOptimizationDecisionDependent2023,NEURIPS2020_33e75ff0,pmlr-v119-perdomo20a}.
Indeed, without any assumptions on the risk measure $\rho$, the cost function $\varphi$ may be non-smooth and non-convex.
However, there are several cases of practical interest where the problem can be solved more efficiently:
\paragraph{Finite action space} Typically, the action space $U$ 
is discretized into multiples of the \textit{smallest tradable unit}, making this set finite and countable, i.e., $U = \{ u_1, \dots, u_n \}$. In this case, Eq. \eqref{eq:solution} can be solved exactly
by evaluating the cost in each $u_i$ and selecting the smallest value.
\paragraph{Continuous cost} Alternatively, if $\varphi$ is continuous
and $U$ is a closed interval, problem \eqref{eq:solution} can be solved with a line-search procedure.
\paragraph{Specific risk measures} For particular choices of the risk measure $\rho$,
even more efficient methods to find the solution may exist. 
For instance, when the expected value is chosen as a risk measure, the optimization problem 
\eqref{eq:solution} can be shown to be both smooth and convex under typical 
conditions (cf. \ref{sec:convex}), which allows the application of more efficient solution methods, such as Newton's method.\\

In the case study (cf. \S\ref{sec:case-study}), the action space $U$ is finite and can be effortlessly enumerated to identify the optimal trade position.

\subsection{Adaptive Risk Parameters}\label{sec:adaptive-strat}
\begin{figure}[!t]
  \centering
  \subfloat[\small $\CVAR$]{\includegraphics[width=88mm]{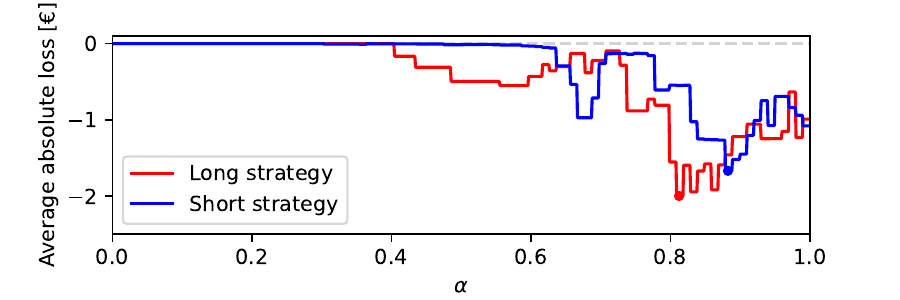}}\\
  \subfloat[\small $\EVAR$]{\includegraphics[width=88mm]{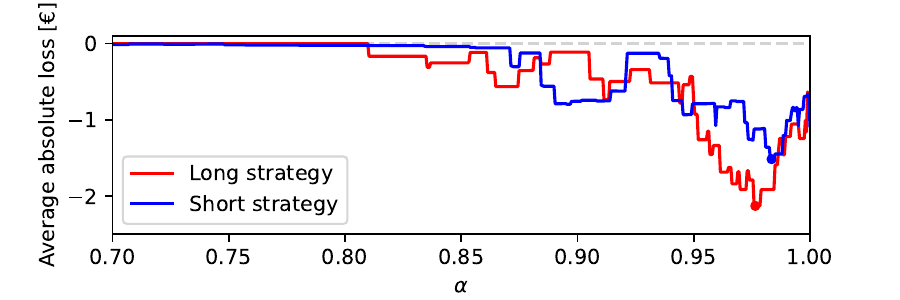}}
  \caption{The average historical absolute loss of the strategy over the past $N=500$ trading periods for different values of $\alpha$ (cost function of Eq. \eqref{eq:opt-alpha}) on the test set of the case study on 2023-06-26 18:30. Negative values indicate profits. For both strategies, the naive approach ($\alpha=1$) results in suboptimal returns. The adaptive trading strategy chooses $\alpha_{t+1}$ as the minimizer of this function.}
  \label{fig:opt-alpha}
\end{figure}

Risk measures introduce a tradeoff between maximizing potential profits and minimizing associated risks. For $\CVAR$ and $\EVAR$, this tradeoff is controlled by the tuning parameter $\alpha$. In very risk-averse scenarios (e.g., safety in autonomous driving \cite{schuurmans_SafeLearningbasedMPC_2023}), a low, constant $\alpha$ is chosen to have high robustness. 
However, in trading, this would lead to very conservative behaviour where the strategy does not take any positions at all. 
Instead, we may afford to take larger risks at times when we are more confident about our predictions. 
To this end, we present a methodology for tuning the risk parameter $\alpha$ based on a window of historical trades.
Concretely, we choose the next $\alpha_{t+1}$ as the value that would have resulted in the  
smallest absolute loss (in hindsight) over the last $N$ trades:
\begin{equation}\label{eq:opt-alpha}
    \alpha_{t+1} = \argmin_{\alpha \in [0,1]} \frac{1}{N} \sum_{i=0}^{N-1} \Big(q_{t-i}\left(u_{\rho_\alpha,t-i}^*\right) - \pSI_{t-i} \Big)u_{\rho_\alpha,t-i}^*
\end{equation}
Here, $\pSI_t$ equals the actual system imbalance price at time $t$, $q_{t}(u)$ the intraday long price and $u_{\rho_\alpha,t}^*$ the risk-averse action that would have been chosen at time $t$ according to Eq. \eqref{eq:risk_averse} with risk measure $\rho_\alpha$.
Using this methodology, the trading strategy will converge to a more risk-averse choice for $\alpha$
when it has consistently generated losses over the past $N$ time steps.
Conversely, a more risk-seeking $\alpha$ will be chosen when consistent profits were made. This cost function, however, is completely non-convex in $\alpha$ as depicted in Fig. \ref{fig:opt-alpha}. 
Evaluating the cost in a single $\alpha$ requires the computation of $N$ optimal trade positions, making it computationally expensive. However, since the optimization problem is solved on a rolling window of historical trades, previously computed solutions can be reused to significantly speed up subsequent evaluations. By discretizing the domain into a fine-grained grid and computing the optimal trade position at each grid point at every timestep, a sufficiently accurate solution can be obtained while also minimizing computational overhead.
Although we have chosen to minimize absolute losses in this paper, this methodology can easily be adapted to other performance metrics.

\section{Case Study}\label{sec:case-study}
In this section, the predictions generated by the previously developed mixture model are used to simulate trading positions on the Belgian electricity market. The strategy is designed to take quarter-hourly positions on the Belgian intraday electricity market five minutes before the cross-border intraday market closes (which is exactly one hour before delivery). To simulate these positions, real-time order books are used from the Nordpool intraday exchange.
The smallest allowed trade unit on the intraday exchange is 0.1MW and a maximum volume of \SI{5}{\mega\watt} is a realistic volume for a small energy trader in Belgium. Hence, the action space is set to $U =  \left\lbrace 0 \mathrm{MW},0.1 \mathrm{MW}, \dots, 5 \mathrm{MW} \right\rbrace$.  The market influence, as outlined in \S\ref{sec:market-impact}, is used to calculate the modified imbalance prices, which are then employed for both training the model and for the profit calculation in the simulation.
A two-year training set spanning from May 1, 2021 to May 1, 2023 is selected to provide the model with sufficient data, while also ensuring that it remains up to date with recent market conditions. Testing is conducted from May 1, 2023 to May 1, 2024. To mitigate seasonal biases, both the training and test sets are chosen as multiples of a year. With renewable energy production increasing annually, market conditions evolve, making larger training sets less representative of current trends.
The inputs to train this model are obtained from either the website of the Belgian TSO Elia or the intraday exchange Nordpool. In this section, we begin by training the statistical model, which is then benchmarked against several models. Finally, the results of the adaptive trading strategy are presented.

  
  

\subsection{Statistical Model}
\begin{figure}
  \centering
  \includegraphics[width=88mm]{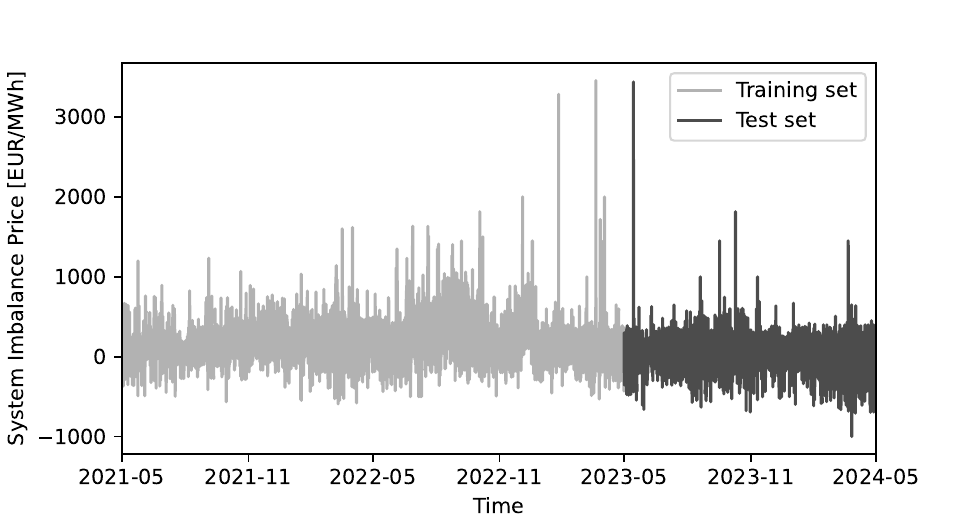}
  \caption{The imbalance prices for both the training and test sets show no clear trends, indicating that the test set serves as a representative sample of the training data. This dataset includes several significant price spikes, primarily caused by nuclear power outages. These spikes were intentionally kept in the dataset to ensure that the model accounts for these extreme price events.}
  \label{fig:imbalance-prices}
\end{figure}

\begin{figure}
  \centering
  \includegraphics[width=88mm]{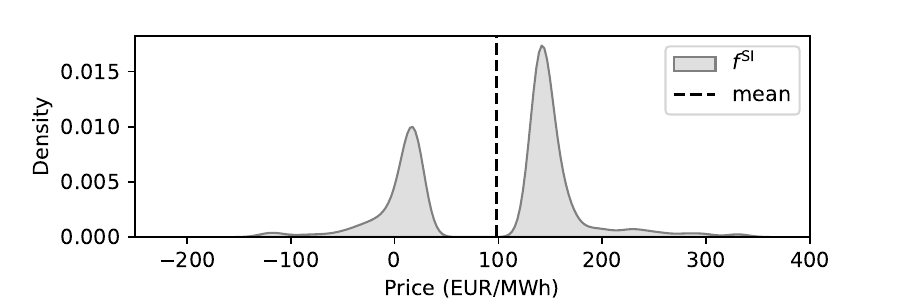}
  \caption{The predicted system imbalance price probability density function $\fSI_t$ on 2023-06-26 18:30 (visualized using kernel density estimation). Because of a higher probability on an energy shortage, the density of the upregulation prices is increased by the mixture weight.}
  \label{fig:distributions_combined}
\end{figure}

In Fig. \ref{fig:imbalance-prices}, the imbalance prices of the test and training set are visualized. There are no clear price trends, making the training set a good representation of the test set.
To model the Belgian imbalance price, we employ the mixture model as proposed in \S\ref{sec:models}. For the mixture weight model, the inputs are the same as those proposed by Koch's imbalance model \cite{kochIntradayImbalanceOptimization2022}. The latter was built for the German electricity market which is very similar to the Belgian one. These inputs include historical system imbalance volumes from the last hour (until one hour before delivery), the quarter hour of the day (one-hot encoded into 96 binary variables), the difference between the intraday and dayahead forecast of the national solar, wind and load, the deviation of the intraday forecast of solar, wind and load from the hourly mean and the difference between the intraday and dayahead price. \\
For the regulation price model, we have to choose the inputs $\ve{z}_t$ and fixed sets of reserve volumes $V_{\Rtwo}$ and $V_{\Rthree}$.
In our experiments, the fixed sets of reserve volumes are 
\begin{align*}
  V_{\Rtwo} &= \{ \SI{1}{\mega\watt}, \SI{50}{\mega\watt}, \SI{100}{\mega\watt}, \SI{150}{\mega\watt}, \SI{200}{\mega\watt} \}\\
  V_{\Rthree} &= \{ \SI{1}{\mega\watt}, \SI{100}{\mega\watt}, \SI{200}{\mega\watt}, \SI{300}{\mega\watt}, \SI{500}{\mega\watt}, \\ &\qquad \SI{700}{\mega\watt} \}
\end{align*}
These volumes are chosen because Elia typically reserves 150MW of aFRR capacity and 700MW of mFRR capacity for the next day \cite{elia_balancing}. Hence, this discretization serves as a realistic approximation of the actual range of reserve volumes.
We choose the output of the regular mixture weight model as the input feature 
$(\ve{z}_t)_{t \in \N}$ of the regulation price model 
(cf. \S\ref{subsec:price-model}).
This value summarizes the market and is strongly related with the amount of reserves that will be activated. For instance, if the probability on an energy shortage is very large, 
then there is also a high probability that the shortage itself will be large and that many reserves have to be activated.
To prevent output data leaking from the mixture weight model to the regulation price model, we apply k-fold crossvalidation to prepare the input for the price model. For both regulation prices, we train 100 different models to predict $n_q=100$ evenly distributed quantiles of that regulation price.
The authors emphasize the importance of selecting a sufficiently large number of quantile models to minimize discretization errors and accurately represent extreme price events in the predicted distribution. A total of 100 quantiles was chosen as an optimal balance between computational efficiency and accuracy. Increasing the number of quantiles beyond this threshold does not yield any differences in performance. 
An example of the forecasted regulation price distributions weighted with the mixture weight is visualized in Fig. \ref{fig:distributions_combined}.

\subsection{Probabilistic Model Benchmark}\label{sec:benchmark}
To evaluate the performance of the previously proposed price model, it is benchmarked against several existing alternatives.
For completeness, we compare against both implicit and explicit models (cf. \S\ref{sec:related-work-model}). Implicit models are designed to predict the system imbalance price directly and in doing so, assume that the model implicitly learns to predict the balancing state. On the other hand, explicit models predict the regulation price distributions separately and then combine them by weighting each distribution with the predicted probability of its corresponding balancing state.
\paragraph{Explicit Benchmark Models}
As mentioned in the introduction, another popular approach for explicitly modelling the imbalance price are regime-switching Markov models (RSMMs) \cite{bunnAnalysisFundamentalPredictability2021,klaeboeBenchmarkingTimeSeries2015,olssonModelingRealTimeBalancing2008,dumasProbabilisticForecastingImbalance2019}.
Markov chains are commonly used to model stochastic processes in which future states depend on the current state.
RSMMs extend this concept by assigning a different predictive model to each Markov state. By weighting the outputs of these predictive models according to their predicted state probabilities, RSMMs are able to forecast prices in the future.
In this study, the Markov states correspond to the two balancing states (i.e. $s_t \ge 0$ and $s_t < 0$) and the price models are chosen to be the same regulation price models as those proposed by our model (i.e. $\fMDP_t$ and $\fMIP_t$).
The key difference between our proposed model and RSMMs lies in how future states are predicted. Our mixture weight model directly estimates the probability to be in a future balancing state, whereas an RSMM predicts each intermediate state sequentially using Markov-chain transition matrices. Since trading positions must be taken at least one hour before delivery, RSMMs are used to predict the balancing state probabilities five quarter-hour intervals in the future. In the benchmark comparison, we discuss two different variants: 
\begin{enumerate}
  \item \textit{Static RSMM:} This RSMM employs a static transition matrix, which is determined by computing the empirical probabilities of transitioning from one balancing state to another in the next timestep based on historical training data.
  \item \textit{Dynamic RSMM:} 
  In reality, the transition probabilities between balancing states are not constant and can be influenced by external factors. To better capture these dynamics, we introduce a more flexible approach where the transition probabilities are conditioned on exogenous input variables (production and load derived variables, quarter of the day as one-hot input, and difference between intraday and dayahead price). To estimate these transition probabilities, we employ two logistic regression models, a method that has been proven successful in previous studies \cite{bunnAnalysisFundamentalPredictability2021}.
\end{enumerate}

\paragraph{Implicit Benchmark Models}
We also present three different non-parametric implicit models: (1) linear quantile regression, (2) artificial neural networks (ANN) with one hidden layer, and (3) random forest quantile regression. Here, all inputs (previous system imbalance volumes, production and load derived variables, quarter of the day as one-hot input, difference between intraday and dayahead price and the order book prices of the fixed set of reserve volumes $\mathbf{o}_t$) are fed into the model at once. 

\begin{figure}
  \centering
  \includegraphics[width=\textwidth]{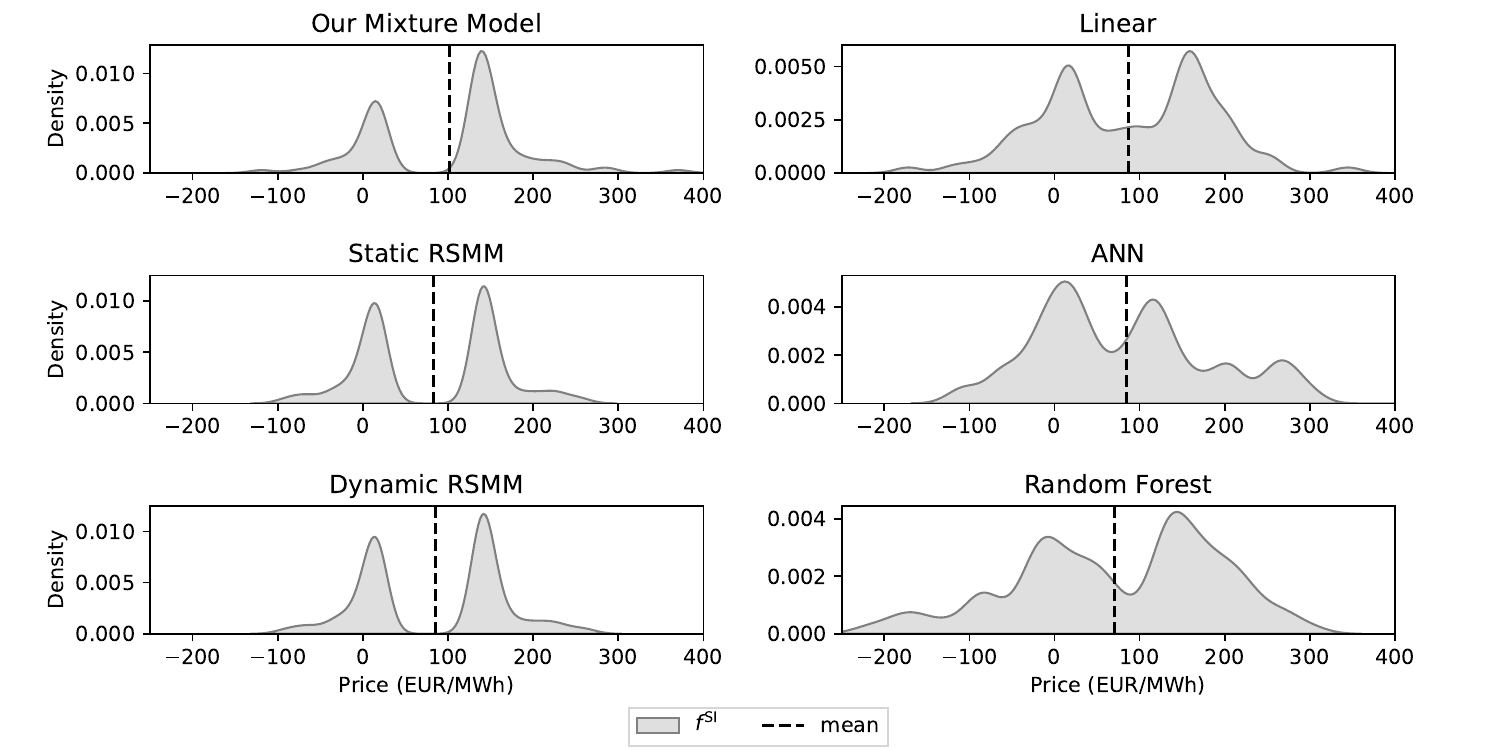}
  \caption{Predicted system imbalance price probability density function of the different models on 2023-05-26 18:30  (visualized using kernel density estimation). The explicit models (left) make a clear distinction between the two regulation prices, while the implicit models (right) predict a less pronounced bimodal distribution. }\label{fig:distribution_forecasted}
\end{figure}

\begin{figure}
    \centering
    \includegraphics[width=\textwidth]{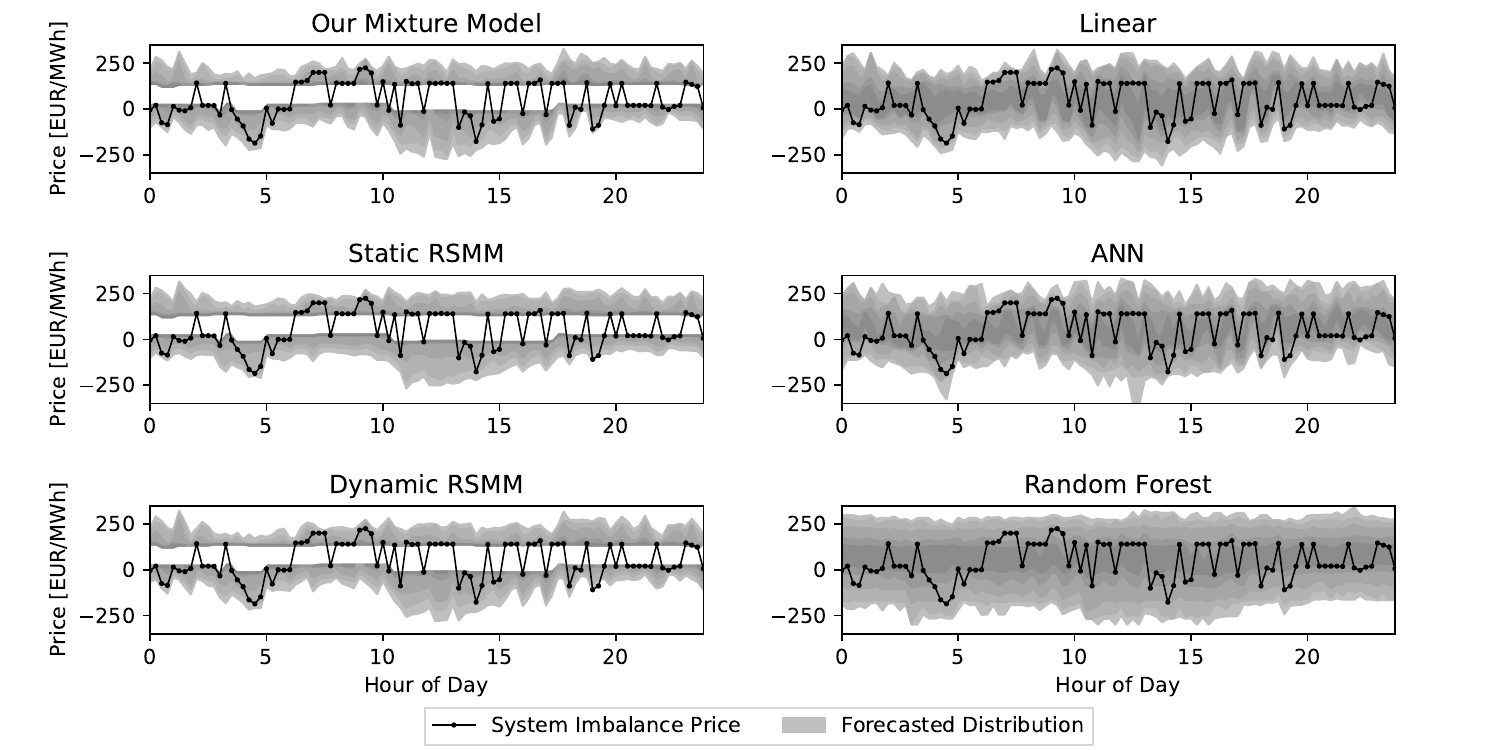}
    \caption{Forecasted distributions on May 26 2023. The alternating behaviour of the system imbalance price (black line) is clearly visible. The explicit models (left) predict denser distributions compared to the implicit models (right). Furthermore, the structural design of the explicit models enables them to account for the gap between the regulation prices.}\label{fig:distribution}
\end{figure}

\begin{enumerate}
  \item \textit{Linear Model:} The linear quantile regression model is the most simple quantile regression model. The output is a linear combination of the input variables. Quantiles are predicted by minimizing the quantile loss as proposed in Eq. (\ref{eq:quantile}). By training separate models for the different quantiles of the system imbalance price, an estimation of its probability density function is obtained. 
  \item \textit{ANN:} Neural network quantile regression models extend traditional linear quantile regression models by replacing the internal model with a neural network. Unlike the linear counterpart, neural network quantile regression can handle complex, nonlinear relationships. Different configurations were tested and a model with one hidden layer and three neurons yielded the best performance. The neurons use the $\mathop{tanh}$ activation function. 
  \item \textit{Random Forest:} Random forest is a popular ensemble learning method for regression and classification. By combining a multitude of decision trees, random forests are able to handle nonlinearity and complex interactions between variables. Meinshausen generalized random forests to not only forecast the mean of an output variable, but also forecast conditional quantiles \cite{meinshausenQuantileRegressionForestsa}. This generalization is also used in this paper. The hyperparameters were chosen by grid search.
\end{enumerate}
Typically, probabilistic forecasts are evaluated on their \textit{reliability} and \textit{sharpness} \cite{gneitingProbabilisticForecasting2014}. 
Reliable models predict quantiles that are statistically correct with their observations, while  
the sharpness refers to the concentration of the predictive distributions. 
More concentrated distributions signify a higher certainty of the model. 
In this study, the sharpness is assessed by calculating the standard deviation (Std) of the predicted distribution. To jointly measure sharpness and reliability, we use the \textit{\ac{crps}} \cite{hersbachDecompositionContinuousRanked2000}, a widely used metric that quantifies the error between the predicted \ac{cdf} and the true \ac{cdf}. 
A lower \ac{crps} equals a better probabilistic forecast. 
We also calculate the \ac{rmse} between the expected value (or mean) of the forecasted distribution and the true observed system imbalance price and the mean absolute error (MAE) between the predicted median of the forecasted distribution and the imbalance price. \\
The performance metrics on the unseen test set are listed in Table \ref{tab:results_forecasts}. 
A snapshot of the forecasted distributions at one timestep is visualized in Fig. \ref{fig:distribution_forecasted} and the probabilistic forecasts during a whole day are depicted in Fig. \ref{fig:distribution}. 
Our proposed mixture model outperforms the other models on every metric. 
In this benchmark, the implicit models struggle to capture the bimodal nature of the imbalance price, leading to wider predicted distributions and relatively higher \ac{crps} scores. 
Despite this, their mean predictions are quite accurate, as reflected in their relatively low RMSE. 
Unsurprisingly, the dynamic RSMM outperforms the simpler static RSMM. 
However, both RSMM variants still fall short of our proposed mixture model. 
Their Markov-based structure limits their ability to accurately forecast balancing states over longer horizons. Having to model all intermediate states increases the risk of accumulating prediction errors. 

\begin{table}[!t]
  
  \centering
  \begin{tabular}{|r|c|c|c|c|}
    \hline
  Model & \ac{rmse} & MAE & Std & \ac{crps} \\\hline\hline
  Our Mixture Model & \textbf{139.18} & \textbf{96.09} & \textbf{108.45} & \textbf{67.33} \\\hline
  Static RSMM & 144.25 & 100.03 & 114.00 & 70.23 \\\hline
  Dynamic RSMM & 141.29 & 98.75 & 110.56 & 68.45 \\\hline\hline
  Linear & 139.36 & 101.75 & 132.45 & 70.24 \\\hline
  ANN & 139.32 & 104.77 & 130.36 & 69.54 \\\hline
  Random Forest & 142.45 & 102.24 & 148.77 & 71.66 \\\hline
  \end{tabular}
  \caption{Losses of different models on the test set. The mixture model outperforms the other models on every metric. The explicit models achieve lower MAE and predict denser distributions. However, these metrics do not necessarily translate into better RSME and CRPS scores. }\label{tab:results_forecasts}
\end{table}

\subsection{Risk-Averse Trading Strategy}\label{sec:trading-results}
\begin{figure}[!t]
  \centering
  \includegraphics[width=88mm]{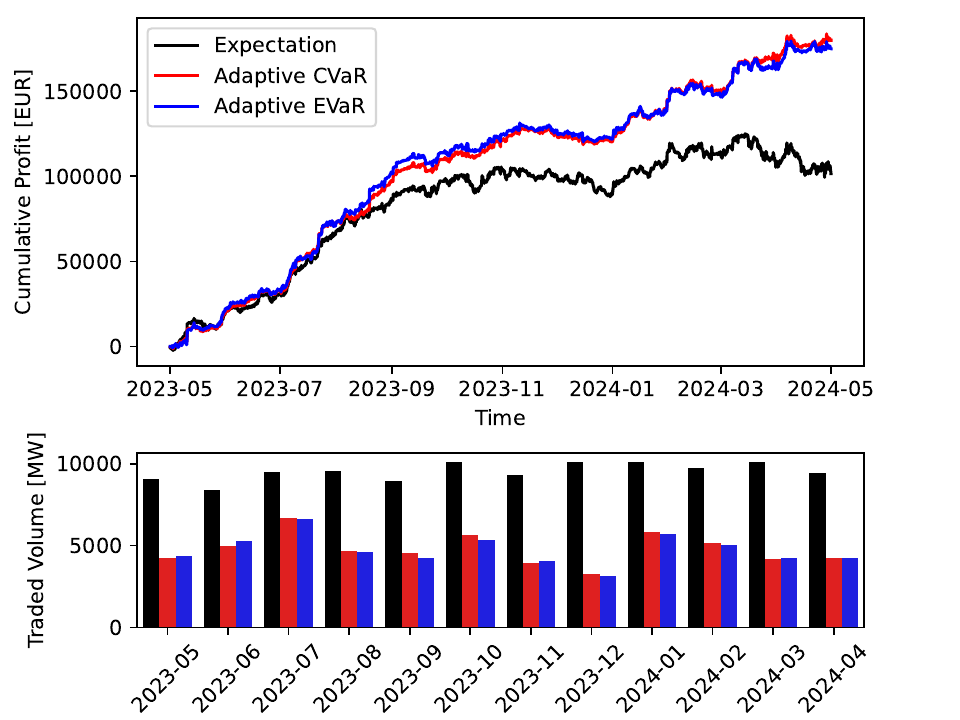}
  \caption{The accumulated profits of the adaptive trading strategies. The adaptive strategies only require a fraction of the trades and generate higher profits.}
  \label{fig:cumulative_profits}
\end{figure}

In this section, the adaptive risk-averse trading strategy is simulated on the Belgian electricity market. To ensure reliability, the order books from the continuous intraday market and inputs for the price model were recorded at the designated trading times, which is always one hour and five minutes before delivery. As mentioned in \S\ref{sec:adaptive-strat}, we compute the optimal trade positions with an enumeration over the 51 volumes in $U$. To calibrate the risk tuning parameter $\alpha_{t+1}$ as outlined in Eq. \eqref{eq:opt-alpha}, a window of $N=500$ previous trades is utilized. The optimal parameter value is chosen by evaluating the expected profits for 200 distinct values of $\alpha$. 
We found that choosing a finer grained grid did not lead to better results.
On an Apple Macbook Pro M1, computing the optimal $\alpha_t$ for a single timestep takes 36ms for $\CVAR$ and 40ms for $\EVAR$. 
This strategy is allowed to take both long and short positions on the electricity market and tracks a separate value for $\alpha_t$ for both position types. In the simulation, we consider the worst-case scenario where the market impact factor $\beta$ from \S\ref{sec:market-impact} is assumed to be 1.\\
The performance of the trading strategy on the test set is detailed in Table \ref{tab:results_strategy} and Fig. \ref{fig:cumulative_profits}. The adaptive trading strategies consistently outperform their static counterparts. 
Moreover, the impact of misestimating the market influence $\beta$ is analyzed by generating imbalance prices with a market impact different from the one assumed by the trading strategy. The findings, summarized in Table \ref{tab:market-impact-adaptive}, highlight that the adaptive trading strategy is relatively robust to misestimations of the market impact.

\begin{table}[!t]
  \centering
  
  \begin{tabular}{|c|c|c|c|c|}
    \hline
    Measure & $\alpha$ & \makecell{Profit \\ \lbrack€\rbrack} & \makecell{Trades \\ \lbrack\si{\mega\watt h}\rbrack} & \makecell{Profit Per \\ Trade [€/\si{\mega\watt h}]} \\\hline\hline
    Expectation & & 101k & 28595 & 3.48  \\\hline\hline
    \multirow{4}{*}{$\CVAR$} & 0.95 &  139k & 22457 & 6.16 \\\cline{2-5} 
    & 0.9 &  154k & 17166 & 8.96  \\\cline{2-5}  
     & 0.8 &  146k & 10002 & 14.55  \\\cline{2-5} 
    & Adaptive & \textbf{171k} & 14402 & 11.87 \\\hline\hline
    \multirow{4}{*}{$\EVAR$} & 0.995 &  149k & 22105 & 6.72  \\\cline{2-5}  
     & 0.98 &  157k & 16328 & 9.60  \\\cline{2-5} 
     & 0.95 &  148k & 9617 & 15.39  \\\cline{2-5}  
     & Adaptive & \textbf{172k} & 14235 & 12.08 \\\hline
  \end{tabular}
  \caption{Performance of the trading strategy on the test set using different risk measures. The adaptive strategies achieve higher absolute profits with fewer trades. As the risk parameter $\alpha$ decreases, the profit per trade increases, while the total number of trades decreases. The adaptive trading strategy automatically searches for the most optimal configuration. }\label{tab:results_strategy}
\end{table}


\begin{table}[!t]
  \centering
  
  \begin{tabular}{|l|c|c|c|c|c|c|}
    \hline
    & $\beta = 0$ & $\beta = 0.2$ & $\beta = 0.4$ & $\beta = 0.6$ & $\beta = 0.8$ & $\beta = 1$ \\\hline\hline
    $\beta_\mathrm{est} = 0$ & 249k &	228k &	212k &	193k & 174k	& 161k \\ \hline
    $\beta_\mathrm{est} = 0.2$ & \textbf{260k} &	\textbf{238k} &	216k & 197k &	180k &	167k  \\ \hline
    $\beta_\mathrm{est} = 0.4$ & 256k &	237k &	\textbf{218k} &	\textbf{199k} &	181k &	170k \\ \hline
    $\beta_\mathrm{est} = 0.6$ & 251k &	232k &	215k &	198k &	178k &	165k \\ \hline
    $\beta_\mathrm{est} = 0.8$ & 246k &	230k &	215k &	194k &	\textbf{182k} &	168k \\ \hline
    $\beta_\mathrm{est} = 1$ & 244k &	229k &	215k &	193k &	178k &	\textbf{171k} \\ \hline
  \end{tabular}
  \caption{Influence of the market impact on the profit of the adaptive $\CVAR$ strategy. Here, $\beta_\mathrm{est}$ represents the estimated market impact (used by the strategy), while $\beta$ denotes the actual market impact (used in the profit calculation). As expected, profit decreases as the actual market impact $\beta$ increases. However, the differences in profit for a fixed $\beta$ and varying $\beta_\mathrm{est}$ remain relatively small. }\label{tab:market-impact-adaptive}
\end{table}

\section{Discussion}\label{sec:discussion} 
The proposed trading strategy is designed to be applied in a real-world setting. The integration of feedback in the adaptive strategy ensures that it remains robust to changing market conditions, which is a crucial element for succesful trading in financial markets.
When comparing the results in Table \ref{tab:results_strategy}, the difference in profits between the adaptive $\EVAR$ and $\CVAR$ trading strategies is relatively small. This outcome aligns with our expectations, as the adaptive risk parameter selection seeks to filter out the same historically unprofitable trades and thereby approximates the same risk ambiguity set. \\
Analyzing the profits in Fig. \ref{fig:cumulative_profits}, it clearly appears more challenging to achieve consistent profitability during the winter months compared to the summer months. This difference may be attributed to the higher levels of renewable energy production in the summer months, making the system imbalance price more predictable with the inputs considered in this study. \\
On May 22, 2024, the Belgian TSO Elia has introduced a deadband mechanism to the imbalance price. Under this framework, the imbalance price is set to a predetermined mid-price when the imbalance volume lies within a specified deadband. To incorporate this deadband into the strategy, adaptations to the model are required to include the new balancing state. 
Furthermore, in November 2024, the Belgian TSO will transition to PICASSO, a platform for sharing reserves among TSOs \cite{PICASSO}. Under this framework, the prices of shared reserves are only disclosed after delivery, in contrast to the local bids of which prices are known in advance. As a consequence, a more sophisticated model will be required to also predict future shared reserve prices. \\
In theory, this strategy may be extended to dual-price markets that include ex-post trading. In these markets, energy contracts are traded after delivery, enabling participants to settle energy imbalances at more favorable prices. Implementing this extension would require a model capable of predicting ex-post market prices to take advantage of this system. A notable example of an ex-post market is the Energy Trading Platform Amsterdam (ETPA), which operates within the Dutch energy market \cite{ETPA}.\\
In addition to the simulation, this exact strategy has recently been implemented by a Belgian BRP and already shows promising results.

\section{Conclusion}\label{sec:conclusion}
Predicting future system imbalance prices is challenging due to their inherently alternating behavior.
In this study, we employed a probabilistic mixture model to effectively capture the underlying distribution of the price, thereby reducing the overall complexity and enhancing robustness against overfitting.
Compared to several benchmark models, the proposed mixture model outperforms its counterparts across every metric.\\
Using this model, we designed a strategy for automated trading on the intraday electricity market. This strategy explicitly accounts for (i) the impact of the trade positions on imbalance price and (ii) uncertainties inherent in the prediction model by including a coherent risk measure in the cost function.
The tuning parameter of this risk measure is adaptively optimized using a window of recently observed data.
After validating this strategy on the Belgian electricity market, the adaptive trading strategy resulted in larger absolute profits with only a fraction of the trades. This outcome highlights the importance of risk awareness in trading strategies to ensure robustness and profitability in real-world applications. 

\section*{Acknowledgements}
This work was supported by Flanders Innovation \& Entrepreneurship (VLAIO) under grant code "Baekeland Mandaat HBC.2022.0695", Research Foundation Flanders (FWO) research projects G081222N, G033822N, and
G0A0920N, and Research Council KUL grant C14/24/103. We gratefully acknowledge Smart@Energy and Amplifino for providing access to the Nordpool intraday energy market data.

\section*{CRediT authorship contribution statement}
\textbf{Robin Bruneel:} Writing - original draft, Validation, Software, Methodology, Investigation, Formal analysis, Data curation, Conceptualization, Visualization, Funding acquisition. \textbf{Mathijs Schuurmans:} Writing - review \& editing, Conceptualization, Methodology. \textbf{Panagiotis Patrinos:} Supervision, Conceptualization, Methodology.

\section*{Declaration of Competing Interest}
Robin Bruneel reports financial support was provided by Amplifino. If there are other authors, they declare that they have no known competing financial interests or personal relationships that could have appeared to influence the work reported in this paper.

\section*{Data Availability}
This study utilizes two primary data sources: Elia and Nord Pool. The data from Elia is publicly available and obtainable from \url{opendata.elia.be}. The Nord Pool data is subject to licensing restrictions and researchers interested in accessing this data must obtain the necessary permissions directly from Nord Pool (\url{www.nordpoolgroup.com}).

\appendix
\section{Efficient Optimization of \eqref{eq:solution} for the Expected Value}\label{sec:convex}
Before discussing more efficient optimization techniques, we begin by proving smoothness and convexity of the cost function $\varphi$ when the risk measure equals the expected value. These are two necessary conditions to use gradient-based optimization methods.

\begin{propositionrestated}
  Suppose $\rho=\E$, and the following conditions hold: 
  \begin{enumerate}[label=(\roman*)]
      \item The regulation price sensitivities are equal: $K_\MDP = K_\MIP = K$\label{cond:K}
  \item The intraday order book within $U$ consists of a single bid such that the intraday energy cost is a linear function of $u$: 
      $$ q_t(u)u = au + b,
      \; a \in \R_+,
      \; b \in \R
      \text{ for all } u \in U. $$
  \item The maximum trade position is sufficiently small: $$u_\mathrm{max} \le\frac{20 K n_q}{\beta w_u^2 \sum_{i=1}^{n_q} \left(g_{i,t}^\MIP - g_{i,t}^\MDP \right)}$$
  \end{enumerate}
  Then, optimization problem \eqref{eq:solution} is smooth and convex.
\end{propositionrestated}
\begin{proof}

Combining condition \ref{cond:K} with Eq. \eqref{eq:modified-si-distribution},
we may write
\begin{equation*}
    \begin{aligned}
        \mathbb{E}_{p \sim \fSI_{t,u}}[-p] &= K\beta u + \pi_t(u) 
        \underbrace{\tfrac{1}{n_q} \sum_{i=1}^{n_q} -g_{i,t}^\MDP}_{c_t^\MDP} +
        (1 - \pi_t(u)) \underbrace{\tfrac{1}{n_q} \sum_{i=1}^{n_q} -g_{i,t}^\MIP}_{c_t^\MIP}, \\
    \end{aligned}
\end{equation*}
and thus, the cost $\varphi$ in \eqref{eq:solution} is given as 
\[ 
    \varphi(u) = \Big( K \beta u + (c_t^\MDP - c_t^\MIP) \pi_t(u) + c_t^\MIP + a\Big) u + b.
\]
This function is twice differentiable:
\begin{equation}
  \begin{aligned}
    \odv{\varphi(u)}{u} &= 2 K \beta u + (c_t^\MDP - c_t^\MIP) \left( \pi_t(u) + \odv{\pi_t(u)}{u}u \right) + c_t^\MIP + a \\
    \odv[order=2]{\varphi(u)}{u} &= 2 K \beta + (c_t^\MDP - c_t^\MIP) \left( 2 \odv{\pi_t(u)}{u} + \odv[order=2]{\pi_t(u)}{u}u \right) \\
  \end{aligned}
\end{equation}
where the derivative of the mixture weight $\pi_t(u)$, being a logistic regression model, is given by:
\begin{equation}
  \begin{aligned}
    \odv{\pi_t(u)}{u} &= \pi_t(u) (1 - \pi_t(u)) \beta w_u \\
    \odv[order=2]{\pi_t(u)}{u} &= \pi_t(u) (1 - \pi_t(u)) (1 - 2 \pi_t(u)) \beta^2 w_u^2 \\
  \end{aligned}
\end{equation}
Here, $w_u$ represents the weight of $u$ in the logistic regression model. From the training set, we obtain $w_u = 0.007$. The derivatives of the cost function $\varphi$ are continuous and bounded within $U$, which implies that the cost function is smooth. 
For convexity, it now suffices to demonstrate that the second-order derivative is positive within $U \subset \R_+$:
\begin{equation}\label{eq:ineq-bound}
  \begin{aligned}
    0 &\le \odv[order=2]{\varphi(u)}{u} \\
    0 &\le 2 K \beta + (c_t^\MDP - c_t^\MIP) \left( 2 \odv{\pi_t(u)}{u} + \odv[order=2]{\pi_t(u)}{u}u \right) \\
  \end{aligned}
\end{equation}
To solve this inequality, we derive a worst-case bound:
\begin{itemize}
  \item The difference between predicted quantiles is always positive. The largest observed difference on the training set is: $\max_t \left( c_t^\MDP - c_t^\MIP \right) = 700$
  \item The sigmoid function, as a function of $u$, always has a positive slope. Thus, the worst-case scenario for this inequality is: $\min_u \odv{\pi_t(u)}{u} = 0$
  \item The minimum second-order derivative of the sigmoid function with respect to $u$ is approximately: $\min_u \odv[order=2]{\pi_t(u)}{u} \approx -0.1 \beta^2 w_u^2$
\end{itemize}
Under these worst-case conditions, inequality \eqref{eq:ineq-bound} simplifies to:
\begin{equation}
  \begin{aligned}\label{eq:ineq-bound-simplified}
    0 &\le 2 K \beta -0.1 \beta^2 w_u^2 u \left( c_t^\MDP - c_t^\MIP \right) \\
    u &\le\frac{20 K}{\beta w_u^2 \left( c_t^\MDP - c_t^\MIP \right)} \\
    u &\le 233 \mathrm{MW}
  \end{aligned}
\end{equation}
As a result, the trade position would need to exceed at least 233 MW for the cost function to become concave. This is an unrealistically large volume for a small energy trader in Belgium, where typical trade sizes are approximately 5 MW. Eq. \eqref{eq:ineq-bound-simplified} is a sufficient condition for the cost function to be convex over its domain.
\end{proof}

The first assumption in the proposition is strictly not necessary, but holds in practice and significantly simplifies the computations in the proof. The second assumption ensures smoothness, as the intraday energy cost $q_t(u)u$ is a piecewise linear function when multiple order book bids have to be filled. In that specific case, the action space $U$ can be subdivided into several segments, within each of which the intraday energy cost remains linear. By applying the proposition to each individual segment, the combined optimization problem becomes both piecewise smooth and piecewise convex. Concavity and non-smoothness may only occur at the (known) hinge points, allowing the problem to be solved efficiently nonetheless.\\
Our empirical findings indicate that the assumptions of the proposition are easily satisfied in practice. The smoothness of the method enables the use of gradient-based optimization techniques or, in this case, where a closed-form expression for the second-order derivative is available, Newton's method for efficiently solving the optimization problem. The convexity guarantees that any stationary point found is also a global minimizer.

\bibliographystyle{elsarticle-num} 
\bibliography{references}
\end{document}